\documentstyle[12pt,preprint,aps]{revtex}

\begin{document}
\title
{Reflectionless tunneling in planar Nb/GaAs hybrid junctions}
\author{Francesco Giazotto$^{a)}$, Marco Cecchini, Pasqualantonio Pingue,
and Fabio Beltram}
\address{Scuola Normale Superiore and Istituto Nazionale per la Fisica 
della Materia, I-56126 Pisa, Italy}
\author {Marco Lazzarino, Daniela Orani$^{b)}$, Silvia Rubini, and Alfonso Franciosi$^{c)}$ }
\address{Laboratorio Nazionale TASC-INFM, Area di Ricerca, Padriciano 99, I-34012 Trieste, Italy}
\footnotetext[1]{Electronic mail: giazotto@cmp.sns.it}
\footnotetext[2]{Also with Institut de Physique Appliquee, Ecole Polytechnique Federale de Lausanne, Switzerland}
\footnotetext[3]{Also with Dipartimento di Fisica, Universit\`a di Trieste, I-34127 Trieste, Italy}
\maketitle


\begin{abstract}
Reflectionless-tunneling was observed in Nb/GaAs superconductor/semiconductor junctions fabricated through a two-step procedure. First, periodic $\delta$-doped layers were grown by molecular beam epitaxy near the GaAs surface, followed by an As cap layer to protect the surface during {\it ex-situ} transfer. Second, Nb was deposited by dc-magnetron sputtering onto the GaAs(001) 2\,$\times$\,4 surface {\it in-situ} after thermal desorption of the cap layer. The magnetotransport behavior of the resulting hybrid junctions was successfully analyzed within the random matrix theory of phase-coherent Andreev transport. The impact of junction morphology on reflectionless tunneling and the applicability of the fabrication technique to the realization of complex superconductor/semiconductor mesoscopic systems are discussed. 
\end{abstract}
\vskip0.5cm

\newpage
Superconductor-semiconductor hybrid junctions have gained considerable attention in recent years because of the new transport phenomena they are bringing to light and in view of their potential technological applications. 
A further important factor in driving the growth of this field is represented by the advancement of nanotechnology and materials science that is opening the way to the experimental observation of novel effects in mesoscopic-scale devices.

Typically hybrid systems consist of coupled superconductor (S) and {\it diffusive} normal-metal (N) (or doped-semiconductor (Sm)) regions. In such configurations electron-electron correlations are induced in the N (or Sm) layers
via a two-electron process known as {\it Andreev reflection} (AR) \cite{andr}. Thanks to this effect, at low bias, an electron originating in the Sm (N) can penetrate in S as a Cooper pair if a hole is (retro)reflected along its time-reversed path with a well-defined phase relationship with the incoming electron. 

One of the most remarkable phase-coherent effects related to AR is {\it reflectionless tunneling} (RT)\cite{marmor}. RT is due to the constructive interference of quasiparticles coherently backscattered towards the interface barrier in the diffusive region\cite{vanwees} and can give rise to a sizable enhancement of conductivity around zero-bias. Its coherent nature makes it very sensitive to electric and magnetic fields and to temperature \cite{kast91,bakk,sanq,magnee}.
 
The key requirement to make possible the observation of AR and of the associated coherent phenomena is the realization of S-Sm
interfaces with good transparency. Achieving the latter in S-Sm contacts, however, is hindered by the presence of interface oxide layers and Schottky barriers, and by the difference in Fermi velocities in the S and Sm layers \cite{btk}. In order to avoid Schottky-barrier formation, several authors choose InAs for the realization of hybrid devices. The main transparency-limiting factor in this case is interface oxidation/contamination and several methods have been explored to limit its impact\cite{kroemer,tak,lach}. InAs, however, is not the Sm of choice for heterostructure growth and GaAs- or InP-based systems are much more widespread and versatile. For these latter materials Schottky barriers are the dominant factor in limiting transparency. In this case penetrating contacts \cite{gao,marsh}, heavily doped surface layers \cite{kast91,tabor} and interface engineering \cite{giaz} were exploited to maximize interface transparency.

In this letter we focus on GaAs-based systems and report the observation of RT in Nb/GaAs junctions. We have developed a versatile technique to fabricate Nb/GaAs hybrid structures with highly reproducible transport properties. Analysis of coherent magneto-transport across the interface was performed within the random matrix theory of Andreev transport. Furthermore the observed dependence of the conductance on magnetic field orientation allowed us to discuss the role of the Nb superconducting mixed state and of interface inhomogeneities in determining RT phenomenology. 

For the semiconducting portion of the structure, a 200-nm thick n-GaAs(001) epilayer (nominal Si doping n\,=\,4.7\,$\times 10^{18}$\,cm$^{-3}$) was initially grown by molecular beam epitaxy (MBE) on an undoped GaAs buffer and a semi-insulating GaAs(001) wafer. This was followed by the growth of 15 nm of GaAs doped by a sequence of five $\delta$ layers (nominal Si concentration $3 \times 10^{13}$ cm$^{-2}$ per layer) spaced by 2.5 nm (see Fig. 1(a)). A similar doping scheme was successfully used by Taboryski et al. to lower the contact resistance in fully-MBE-grown Al/GaAs(001) junctions \cite{tabor}. The choice of {\it in-situ} junction formation, however, severely limits the superconductor materials usable. In our case, on the contrary, following the growth of the top $\delta$-doped GaAs layer, a 1-$\mu$m-thick amorphous As cap layer was deposited at -20$^\circ$C to protect the surface during transfer in air to the sputter-deposition/surface analysis system (base pressure 5\,$\times$\,10$^{-10}$ Torr). The sample was then heated at about 400$^\circ$C to desorb the cap layer and achieve a sharp GaAs(001) 2\,$\times$\,4  reflection high energy electron diffraction pattern \cite{fan}. Nb overlayers 100-nm-thick were fabricated {\it in situ} by dc-magnetron sputter-deposition at a deposition rate of 3.5 nm/s. Substrate temperature during Nb deposition was kept at $\approx$ 200\,$^\circ$C to promote film adhesion. A typical 100-nm-thick Nb film displayed a transition temperature $T_{c}$ of 9.28\,K and a high residual resistivity ratio (RRR) at 10\,K of RRR\,$\approx$\,60 (data not shown). 

Rectangular 100$\times$160\,$\mu$m$^2$ Nb/GaAs junctions were patterned by 
standard photolithographic techniques and reactive ion etching (RIE) using a 
CF$_{4}$+O$_{2}$ gas mixture. Two additional 
90$\times$45\,$\mu$m$^2$ Ti/Au bonding pads were e-beam evaporated on top of 
each Nb contact to allow 4-wire measurements (see Fig. 1(b)). These were performed with two leads on the Nb electrode under interest and the other two connected to two separate neighboring contacts located symmetrically with respect to the junction considered. In this way our conductivity data reflect only the junction properties with no influence from the series resistance of the semiconductor film.
We have also experimentally determined the influence of the leakage paths around the contacts by removing the semiconductor layer above and below the contact strip (``contact-end test structure'' \cite{scho}). The correction to the contact resistance is negligible on the scale relevant to the experiment. 
Magnetoconductance measurements
of the junctions were performed in a closed-cycle $^3$He cryostat 
equipped with a superconducting magnet. In order to 
determine carrier concentration and Hall mobility a portion of the 
semiconductor was not Nb coated and, following As desorption, it was patterned into Hall bars by removing the entire epilayer. At 0.34\,K we obtained $n=3.1\times 10^{18}$\,cm$^{-3}$ 
and $\mu = 1.6\times 10^{3}$\,cm$^{2}$/V s. From these values it was possible 
to estimate electron mean free path (${\ell}_{m}\,\approx 50$\,nm) and  
thermal coherence length at $T$\,=\,0.34\,K: 
$\xi (T)=\sqrt{\hbar D /2\pi k_{B} T}\approx\,0.21\,\mu$m, 
where $D=1.2\times 10^{2}$\,cm$^{2}$/s is the three-dimensional diffusion constant. Additionally the single-particle phase coherence length 
$\ell_{\phi}\alt 1$\,$\mu$m was also estimated from weak localization 
magnetoresistance measurements, following Ref. \cite{kut}.

Figure 2 shows a typical set of differential-conductance $vs$ bias 
characteristics ($G(V)$) of one of the junctions measured at various 
temperatures in the 0.34\,-\,1.00\,K temperature range. The symmetry of the
characteristics together with the large value of the conductivity at
low bias at the lowest temperatures are a clear indication of the 
transparency of the Nb/GaAs junctions fabricated (see the inset of Fig. 2
where the conductance spectrum at 0.34\,K up to biases well above the superconducting energy gap  is plotted). The fabrication procedure adopted yielded an excellent contact-to-contact uniformity leading to $G(V)$ characteristics reproducible within few \% in all the junctions measured. The data display the typical behavior of a diffusive non-tunnel S-Sm junction. In particular, at the lowest temperatures measured, a marked peak is observed in the differential conductance at zero bias. As we shall argue, this is the manifestation of RT and it will be very useful for the analysis of the junction properties. In fact it is not straightforward to obtain reliable junction parameters by studying the Andreev pattern in diffusive junctions \cite{antoin,neurohr}. On the contrary, when one such zero-bias peak (ZBP) is observed in these systems, much information can be obtained by
examining its intensity and its temperature and magnetic-field 
dependence \cite{marmor}. 

The electrical properties of our GaAs/Nb junctions can be analyzed within the random-matrix theory of phase-coherent Andreev reflection proposed by Beenakker \emph{et al.} \cite{reviewBEEN}. These authors studied a model-junction consisting of a normal and a superconductor reservoirs linked by a disordered normal-metal region of length $L$ and width $W$. 
Between this disordered region and the superconductor reservoir the model-system includes a potential barrier characterized by a mode-independent transmission probability ($\Gamma$). In this system, under appropriate conditions an enhanced conductivity around zero bias was predicted to occur with respect to the {\it classical} AR behavior leading to a ZBP. The width of the latter (i.e. the bias voltage at which the RT correction to the conductivity vanishes) is given by 
$V_{c}=(\pi /2)\hbar v_{F} \ell_{m}/eL^{2}$, where $v_{F}$ is the Fermi velocity in the normal region. Furthermore, upon application of a magnetic field ZBP is suppressed for field intensities larger than $B_{c}=h/eLW$. In real systems, at finite temperatures, $L$ and $W$ are to be replaced by $\xi(T)$, if it is smaller \cite{marmor}.  

In order to analyze our data within this model, we can first of all determine the experimental value of $V_c$ by examining the data in Fig. 3(a) where the evolution of the ZBP is well displayed thanks to the limited influence of the in-plane magnetic field on the classical conductivity. The RT correction persists up to about $V_c\approx 0.6$\,mV corresponding to a characteristic length $L \approx$\,0.25\,$\mu$m. As predicted in Ref. \cite{marmor}, this value is in good agreement with the estimated thermal coherence length at the same temperature. 

Further insight on the interference effects leading to RT can be gained
observing the magnetic-field dependence of the ZBP.
Figure 3 shows a set of $G(V)$'s at $T=0.34$\,K for several 
magnetic fields applied (a) in the plane of and (b) perpendicularly to the junction. At relatively weak magnetic fields $\sim$\,mT the peak is suppressed and the subgap conductance dip shrinks due to the magnetic-field-induced suppression of the superconductor energy gap. At higher values of the applied magnetic field the overall subgap conductance increases approaching its normal-state value (data are not shown for clarity). From the in-plane-field data, we can determine with good accuracy the experimental critical magnetic field $B_{c} \approx 100$\,mT corresponding to RT suppression and to the minimum measured value of the zero-bias conductivity. 
This value is in good agreement with the expected theoretical value $B_{c} \approx 80$\,mT, obtained from the estimated thermal coherence length $\xi(T=0.34\,\mathrm{K})\approx 0.21\,\mu$m confirming the physical origin of the observed conductance behavior. To the best of our knowledge RT in a GaAs/Nb junction was never reported and was made possible by the fabrication procedure adopted in this work.

It is interesting to compare the conductance behavior caused by different field orientations. For perpendicular fields the overall subgap conductance increases at much smaller fields as compared to the in-plane configuration. We attribute this effect to the type-II nature of Nb and  to the consequent appearance of vortices in the mixed state. The presence of these normal regions within the junction leads to increased subgap conductance  due to the lower resistance of the normal vortex region with respect to the superconducting portions \cite{nota}. We should like to stress the importance of  
careful analysis of the in-plane field curves for a quantitative determination of the RT correction to zero-bias classical conductivity without any empirical extrapolation. 

The quantitative analysis of the ZBP intensity, within the model adopted so far can yield an estimate of junction transmission probability ($\Gamma$)\cite{magnee}. However, the transmissivity calculated from the ratio between the resistance values taking into account the phase-coherent AR and the classical behavior for S-N systems are incompatible with the phenomenology observed. In fact, for our junction properties ($L >> \ell_m$), the observed phenomenology (i.e. $increased$ conductance at zero bias) is expected for $\Gamma \alt 0.4$ while the above procedure leads to significantly higher values. This deviation in the observed intensity of the ZBP must be linked to a significant suppression of the RT-induced enhancement of the conductivity. We believe that this effect stems from junction inhomogeneities. Several authors have reported that only a very small fraction (10$^{-2}$--10$^{-4}$) of the junction physical area determines the transport properties of large diffusive contacts \cite{kast91,gao,giaz,vanhuf}. Additionally, the very good contact-to-contact uniformity we observed indicates that the lateral scale of the higher-transparency regions must occur on an even smaller characteristic length. This peculiar junction morphology leads to a drastic reduction in the number of the reflectionless paths available for ZBP build-up and to the observed suppression of the ZBP intensity.

In conclusion, we have reported on RT-dominated transport in Nb contacts on GaAs and presented an effective and reproducible fabrication technique. The observed regime was studied as a function of temperature and magnetic field. Data were analyzed within the random-matrix theory of phase-coherent AR and the impact of junction non uniformity on RT was underlined. The fabrication technique presented is compatible with the most widespread MBE systems and can be easily exploited for the implementation of novel hybrid S-Sm mesoscopic systems by tailoring the band-gap profile of the AlGaAs/GaAs heterostructures on which the Nb film is deposited and patterned. 

The present work was supported by INFM under the PAIS project Eterostrutture Ibride Semiconduttore-Superconduttore. One of the authors (F.G.) would like to acknowledge Europa Metalli S.p.A. for financial support. One of the authors (M.C.) would like to acknowledge the Consiglio Nazionale delle Ricerche for financial support. 


\begin{figure}
\caption{ (a) Schematic cross section of the Nb/GaAs junctions studied in this work. (b) Scanning electron microscope image of one of the devices.
}
\label{F1}
\end{figure}

\begin{figure}
\caption{Differential conductance $vs$ voltage at several temperatures in the 0.34 to 1\,K range. The inset shows the conductance at $T=0.34$\,K in a wider bias range. The reflectionless tunneling enhancement of the conductivity appears as a peak at around zero bias.
} 
\label{F2}
\end{figure}

\begin{figure}
\caption{ Differential conductance $vs$ voltage at $T=0.34$\,K for several magnetic field values. (a) Magnetic field applied in the plane of the junctions; (b) magnetic field applied perpendicularly to the junctions. 
}
\label{F3}
\end{figure}


\begin{references}
\bibitem{nota} A quantitative estimate of the normal-zone contribution to the conductivity is hindered by the uncertainty about the effective junction area contributing to the transport (see text).
\end{references}
\end{document}